\documentclass[conference]{IEEEtran}
\IEEEoverridecommandlockouts
 
\usepackage{cite}
\usepackage{amsmath,amssymb,amsfonts}
\usepackage{graphicx}
\usepackage{cuted}
\usepackage{capt-of}
\usepackage{placeins}
\usepackage{textcomp}
\usepackage{xcolor}
\usepackage{booktabs}
\usepackage{url}
\def\BibTeX{{\rm B\kern-.05em{\sc i\kern-.025em b}\kern-.08em
    T\kern-.1667em\lower.7ex\hbox{E}\kern-.125emX}}

\begin{document}

\title{Isothermal--Isobaric Molecular Dynamics with a Foundation Interatomic Potential in LAMMPS across NVIDIA and Intel GPUs}

\author{\IEEEauthorblockN{Xiaoli Yan, Eliu Huerta}
\IEEEauthorblockA{\textit{Argonne National Laboratory} \\
Lemont, IL, United States \\
xyan11@anl.gov, elihu@anl.gov}
}

\maketitle

\begin{abstract}
Foundation machine-learning interatomic potentials (MLIPs) offer one model for many chemistries, but their deep, globally coupled architectures do not fit the spatial decomposition that scales classical and short-ranged MLIPs. We present \texttt{pair\_style uma}, a LAMMPS pair style that evaluates the Universal Model for Atoms (UMA) through LibTorch on NVIDIA and Intel GPUs from a single source tree. It reimplements FairChem's edge-sharded graph parallelism in C++ over exported TorchScript modules, with one MPI rank per device, NCCL or oneCCL collectives, and the whole energy and force path in double precision. The strain-gradient virial is carried through the exported modules and the cross-device reductions, so that \texttt{fix npt} runs on both vendors' hardware.

We measure NPT capacity and cost on one 4$\times$A100-40GB node, one 8$\times$H200 node, and one 12-tile Intel node, and on up to 32 Intel nodes. Without activation checkpointing, one node holds NaCl cells of about 8{,}000 atoms (A100, 0.21~ns/day), 22{,}000 atoms (Intel, 0.10~ns/day), and 47{,}000 atoms (H200, 0.16~ns/day). Graph parallelism raises the capacity of a single device by $8.0\times$ on H200 and $4.6\times$ on A100. Across nodes, capacity grows to about 779{,}000 atoms on 32 Intel nodes, but speed does not: the rate falls to 0.007~ns/day, because extra nodes add memory rather than shorten the step. The virial adds about 0.2\% per step over NVT. On H200 and A100, a FLOP count of FairChem's graph-parallel predictor gives lower bounds of 31\% and 51\% of FP64 peak.

Forces agree with an ASE FairChem reference to $\sim\!10^{-14}$~eV/\AA, and thermodynamic output agrees across all device counts that fit on each GPU type; FairChem's \texttt{fix external} wrapper, which passes single-precision coordinates, deviates by $10^{-6}$--$10^{-5}$~eV/\AA. From one \texttt{omat} checkpoint and with no refit, nine 100~ps NPT heating ramps (NaCl, equiatomic Pt--Ru, and equiatomic (HfNbTaTiZr)N, on all three node types) ran to completion.
\end{abstract}


\section{Introduction}

Foundation machine-learning interatomic potentials (MLIPs) trained on hundreds of millions of structures
\cite{wood2025uma} occupy a different point in the parallel design space
from the short-ranged potentials that weak-scale to a full machine.
A single Universal Model for Atoms (UMA) checkpoint, with task heads for materials (\texttt{omat}),
catalysis (\texttt{oc20}), etc., models
chemistries that lack a transferable classical interatomic potential; that transferrability costs the locality multi-node parallelism needs. UMA stacks four
message-passing blocks at a 6~\AA\ cutoff \cite{wood2025uma,fairchem2025}, a $\sim$24~\AA\ receptive field,
and two global couplings: channel balancing \cite{fairchem2025} and the
mixture-of-linear-experts (MoLE) routing \cite{wood2025uma}. A correct spatial halo is four shells
deep, not one pair cutoff. At the cell sizes this ensemble can hold, even one
shell is comparable to the owned subdomain (Sec.~\ref{sec:internode}), so a
spatial split leaves the halo as large as the work it surrounds. What
remains correct without model surgery is graph parallelism, which keeps the
configuration global and shards only the edges. Every block then gathers the
full node-feature tensor, so communication is $O(N)$ per layer and
per-device node memory is $O(N)$ independent of world size. Both properties are acceptable inside a node, but they tend to slow down performance when internode communication happens: more nodes no longer means better simulation performance.
(Sec.~\ref{sec:internode}). One node is therefore the unit of speed for this model class, and further nodes provide capacity. The systems question is how to put that evaluation inside the Large-scale Atomic/Molecular Massively Parallel Simulator (LAMMPS)
\cite{thompson2022lammps} so that microcanonical (NVE), canonical (NVT), and especially isothermal-isobaric (NPT) simulations run on the NVIDIA and Intel data-center GPUs already installed in high-performance computing (HPC) centers.

The distinction that matters in this systems paper is between an
\emph{ad hoc} and a \emph{foundation} potential. The million- and
billion-atom MLIP results that define the current scaling frontier were all
obtained with a potential trained for one system: Spectral Neighbor Analysis
Potential (SNAP)
\cite{thompson2015snap} for carbon at 20
billion atoms \cite{nguyencong2021billion}, Deep Potential for
copper and water at 100 million \cite{jia2020pushing}, Allegro for one
solvated biomolecular assembly at 44 million \cite{kozinsky2023scaling}.
Each is short-ranged and locally decomposable by construction, which is
precisely why it weak-scales to a full machine; the cost is a training
campaign per chemistry. We use \emph{transferable} in the strict sense: a potential is transferable if changing the
chemical elements or composition requires \emph{no new parameter fit or
retraining}. By that measure the established approaches fail: classical embedded-atom
\cite{daw1984eam},
polarizable-ion and fixed-charge force fields, and ad hoc per-system MLIPs
alike, all refit per chemistry. UMA takes the opposite position. One checkpoint,
selected at export to a task head, evaluates forces across molecules,
materials and catalysis, and the released model matches or beats specialized
models without fine-tuning \cite{wood2025uma}. For the demonstration workload in this
paper (heating ramps of NaCl, equiatomic Pt--Ru, and equiatomic
(HfNbTaTiZr)N from one \texttt{omat} checkpoint), a classical potential would
need three separately parameterized tables for a salt, a binary alloy, and a
five-cation nitride, whereas one foundation model runs all three.

Three constraints then dominate the design.

First, parallelism here pools memory rather than providing speed. Edge sharding
across the devices of one node raises the largest cell that fits.

A second node does not shorten the step: per-rank node-feature work stays
$O(N)$, and the node-boundary cost itself is $\sim$2\% of the NPT step
(Sec.~\ref{sec:internode}). Capacity grows by about 23 thousand atoms per node, but the step time grows with the cell. The
remaining multi-node axis is data parallelism over independent configurations.

Second, NPT runs with checkpointing off. The barostat needs a strain gradient, which lets the cell change volume through a transition, and we have not validated that gradient through our activation-checkpointing path (Sec.~\ref{sec:ckpt}). Running with checkpointing off costs capacity: the checkpoint-on NVT ceiling on the same node is 20 times higher.

Third, portability is a requirement rather than a feature. Centers are deploying both NVIDIA and Intel accelerators, and a molecular dynamics (MD) engine whose results depend on which vendor's hardware runs it is not suitable for scientific use. We therefore treat
cross-device agreement as a correctness property to be measured.

Graph parallelism and the autograd virial are FairChem's (Sec.~\ref{sec:gp}). Our contribution is a new capability, an FP64, NPT-capable LibTorch pair style that also runs on Intel GPUs, not a speedup over an existing solution. It comprises three contributions:

\begin{enumerate}
  \item \textbf{A LibTorch interface from LAMMPS to UMA.} A TorchScript export and a C++ engine behind \texttt{pair\_style uma}, with one MPI rank per device and FP64 end to end, built from one source tree for NVIDIA GPUs (NCCL collectives) and Intel GPUs (oneCCL collectives, named XCCL in PyTorch). Thermodynamic output at steps 0 and 10 agrees across every device count that fits on each GPU type, and forces match an ASE FairChem reference to $\sim\!10^{-14}$~eV/\AA\ (Sec.~\ref{sec:correctness}).
  \item \textbf{The strain-gradient virial through that interface, for NPT.} FairChem's predictor already derives stress from position and cell gradients; we carry it through exported modules, graph parallelism, and \texttt{fix npt}, and show it stays nearly free. On the 22{,}000-atom cell, ten NPT steps take 8.54~s against 8.52~s for NVT, because the virial rides the same backward pass as the forces
  (Sec.~\ref{sec:capacity}).
  \item \textbf{NPT molecular dynamics of up to 47{,}000 atoms at 0.157~ns/day, with delivered 100~ps runs.} We measure the checkpoint-off NPT ceilings on 4$\times$A100-40GB, one 12-tile Intel node, and 8$\times$H200 (Sec.~\ref{sec:capacity}), and the Intel ceiling out to 32 nodes (Sec.~\ref{sec:internode}). On all three node types, one \texttt{omat} checkpoint then runs 100~ps NPT heating ramps of NaCl, equiatomic Pt--Ru, and equiatomic (HfNbTaTiZr)N with no new fit (Sec.~\ref{sec:melt}).
\end{enumerate}

\textbf{Scope.} 
The headline results are node-local by design; Sec.~\ref{sec:internode} adds
one multi-node capacity measurement, to 32 nodes. A spatial-DD path
exists in the source but is not validated and is not claimed; the halo
comparison above is why a thin halo does not apply. We claim a capacity gain only over a single device, the regime of the ASE FairChem calculator, and a precision difference, not a capacity or speed advantage, over FairChem's LAMMPS wrap. We also do not claim nanosecond trajectories (Sec.~\ref{sec:capacity}). The heating-ramp jumps of
Sec.~\ref{sec:melt} are not melting points;
Sec.~\ref{sec:melt-accuracy} states why. Section~\ref{sec:related} places this against
custom million-atom MLIP MD, other vendors' UMA stacks, and Python drivers.

\section{Background}

\subsection{UMA in MD}
UMA \cite{wood2025uma} represents a configuration as a graph with edges
inside a cutoff (6~\AA\ in the exported \texttt{uma-s-1p2} artifacts used
here). Equivariant message-passing layers update node features; energy
is a sum of per-atom terms and forces are its gradient. Task heads
(\texttt{omat}, \texttt{oc20}, \ldots) are selected at export. Mixed-element
systems are legal at runtime; the neighbor graph, not a classical
cross-term file, supplies the multi-element interactions. Two properties
drive everything below: the model is deep, so its receptive field is
$L$ times the cutoff for $L$ blocks ($4\times6 = 24$~\AA\ here), and it
carries global couplings (channel balancing and the mixture-of-linear-experts
routing) that are not expressible as a neighborhood reduction.

\subsection{Why graph parallel, not spatial domain decomposition}\label{sec:gp}
LAMMPS spatial decomposition assigns atoms by box region and ghosts a
halo of width equal to the pair cutoff, and this is what lets custom,
short-ranged MLIPs weak-scale to whole machines
\cite{nguyencong2021billion,jia2020pushing,kozinsky2023scaling}. For a deep
graph neural network (GNN) the physical cutoff is not the receptive field: a correct halo must be
$L$ shells deep, and at $L{=}4$ with a 6~\AA\ cutoff the ghost region can
exceed the owned partition, which we observe directly
($n_\mathrm{all}\approx62$k ghosts per rank for a 74{,}000-atom cell at
$k{=}1$). The global couplings then remain unhandled even with a deep halo.

Graph (edge) parallelism instead keeps the configuration global and shards edge work across devices, using collectives to gather node features. This is the multi-GPU mode of FairChem's own predictor \cite{fairchem2025}: edges are split by index, node features are all-gathered before each block, and forces and the cell gradient are all-reduced. We reimplement it in C++ over exported TorchScript modules behind a LAMMPS pair style, with NCCL and oneCCL collectives. The design is FairChem's; the engine, the Intel GPU port, and the measurements are ours. It is
correct for this architecture without model surgery, and it is the mode in which we have both a working NPT barostat and an implementation that runs on both CUDA and XPU devices. Its
two costs are structural and set the scope of this paper: each block gathers
the full $[N,C]$ node tensor onto every rank, so communication is $O(N)$ per
layer and per-device node memory is $O(N)$ \emph{independent of world size}.
Only edge work is divided. Inside a node, over NVLink or Xe~Link, this is a
good trade; Sec.~\ref{sec:internode} shows a second node provides memory but not speed.

\subsection{Why checkpoint-off NPT}\label{sec:ckpt}
The NPT virial is a derivative through cell strain. Activation checkpointing \cite{chen2016checkpoint} is compatible with it in principle: FairChem's eager model checkpoints its edge-chunked convolutions with \texttt{torch.utils.checkpoint} and still differentiates through the cell. A traced module cannot use that API, so our engine recomputes chunks and blocks in custom C++ autograd functions, and we have not validated the cell gradient through them. The pair style therefore refuses the combination, and every NPT result below runs with checkpointing off (\texttt{UMA\_AC=off}, \texttt{UMA\_CKPT} unset); if either switch is on, the pair style aborts \texttt{fix npt}. The cost below is a limit of our implementation, not of the physics. The price is capacity: on one 12-tile Intel node the checkpoint-on NVT ceiling is $N{=}38$ (439{,}000 atoms, 28.8~GiB peak), 20 times the 22{,}000 atoms of the checkpoint-off NPT ceiling.

\section{Related Work}\label{sec:related}

We do not claim the largest MLIP MD. We claim that the
\emph{foundation}-MLIP regime is a different systems problem from the
custom-MLIP regime that defines the scaling frontier, and that in this regime
a correct barostat and cross-device agreement, not atom count, are the
binding constraints.

\subsection{Scale frontier and foundation models}
The record-setting MLIP MD results are all single-chemistry: Nguyen-Cong~et~al.\ \cite{nguyencong2021billion} run 20~billion carbon atoms at 50.0~peta floating-point operations per second (PFLOPS) on Summit; Jia~et~al.\ \cite{jia2020pushing} reach 100~million atoms at 91~PFLOPS with Deep Potential; Kozinsky~et~al.\ \cite{kozinsky2023scaling} run Allegro \cite{musaelian2023allegro} on a 44-million-atom solvated HIV capsid and report 70\% weak scaling to 5120~A100s. Each uses a potential trained for its target system, with a short receptive field that decomposes spatially with a thin halo.
UMA's $\sim$24~\AA\ receptive field and global couplings force graph parallelism, which pools memory within a node and caps the cell at tens of thousands of atoms per node (Sec.~\ref{sec:internode}). In exchange, one checkpoint spans molecules, materials, and catalysis without fine-tuning \cite{wood2025uma}: NaCl, equiatomic Pt--Ru, and an equiatomic five-cation nitride from one binary with no retraining, while no single transferable classical potential spans that set. Papers at the scale frontier report NVE/NVT; we know of no published strain-gradient virial NPT result at foundation-MLIP cost.

\subsection{Integration and baseline context}
Deep Potential \cite{jia2020pushing,wang2018deepmd}, SNAP \cite{thompson2015snap,nguyencong2021billion}, Allegro/Neural Equivariant Interatomic Potentials (NequIP) \cite{musaelian2023allegro,batzner2022nequip,kozinsky2023scaling}, and MACE \cite{batatia2022mace} are delivered as LAMMPS pair styles; the integration plumbing is established prior art. What changes at foundation scale is the parallel decomposition, memory envelope, and virial path. NVIDIA ALCHEMI \cite{alchemi} runs UMA across several NVIDIA GPUs from a Python driver on the Warp/CUDA stack; to our knowledge it has no Intel XPU path, and this paper does not compare against it.
Other MLIP families reach LAMMPS through spatial decomposition: SevenNet \cite{park2024sevennet} exchanges node features between message-passing layers for NequIP-based models, and chemtrain-deploy \cite{fuchs2025chemtrain} deploys semilocal MACE, Allegro, and PaiNN potentials through LAMMPS domain decomposition. Both build on message passing with a short per-layer reach and are established options for large cells. A UMA user trades speed per atom for one checkpoint spanning molecules, materials, and catalysis; this paper makes that choice available as an FP64, NPT-capable pair style on NVIDIA and Intel GPUs, and does not claim it is faster than those engines. For context, SevenNet reports over 80\% weak-scaling efficiency on 32 GPUs \cite{park2024sevennet} and chemtrain-deploy scales to millions of atoms \cite{fuchs2025chemtrain}; the models and hardware differ, so these are not comparisons. To our knowledge no systematic measurement of multi-device UMA capacity ceilings and NPT step cost has been published. The UMA paper reports that UMA-S fits 100{,}000 or more atoms on one 80~GB GPU in its default inference settings \cite{wood2025uma}. Our NPT ceilings are lower because they are FP64, differentiate through the cell, and retain activations (checkpointing off); the checkpoint-on NVT ceiling of 439{,}000 atoms on one Intel node (Sec.~\ref{sec:ckpt}) is the closer comparison.
The Atomic Simulation Environment (ASE) \cite{larsen2017ase} and FairChem \cite{fairchem2025} Python drivers are the quantitative baseline throughout: identical geometry, tiles, and FP64 model. FairChem also ships a LAMMPS path (\texttt{fairchem-lammps}) that drives LAMMPS through \texttt{fix external} rather than a pair style \cite{fairchemlammps,thompson2022lammps}; that wrap is a second baseline in Sec.~\ref{sec:correctness}, not the engine under test. The Python driver on that geometry recomputes the stress under a different integrator (Nos\'{e}--Hoover here \cite{nose1984,hoover1985,shinoda2004npt}, Berendsen-style in the driver \cite{berendsen1984}), so the timing in Sec.~\ref{sec:capacity} is cost per step. FairChem's LAMMPS wrap is the closer baseline, and it is not slower than \texttt{pair\_style uma}. Classical embedded-atom method (EAM) \cite{daw1984eam,zhou2004eam} and modified embedded-atom method (MEAM) \cite{baskes1992meam} are fit per element set; one \texttt{omat} checkpoint evaluates NaCl, Pt--Ru, and (HfNbTaTiZr)N with no new fit, and a classical step on any one table is faster (this paper carries no speed comparison).

\section{Design}

\subsection{Pair style}
\texttt{pair\_style uma} builds the neighbor graph from LAMMPS neighbor
lists, runs a traced PyTorch C++ (LibTorch) module, and returns per-atom energies and
forces. \texttt{newton off}, \texttt{units metal}, timestep $0.001$~ps
are the production conventions. Export is offline: a geometry
(\texttt{structure.xyz}) is traced once per $(N,W,\mathrm{task})$
triple because \texttt{edge\_pad\_cap} is baked from that graph.
NaCl artifacts are not reused for the alloy cell (different coordination
at the same cutoff), nor a smoke export for a 20k-atom cell.

\subsection{Launch contract}
The launch contract is enforced by the pair style, not merely followed by
convention:

\begin{itemize}
  \item $W$ Message Passing Interface (MPI) ranks \cite{mpi40} on the $W$ devices of one node, or across nodes in the multi-node runs, rank $r$ bound to
  device $r$; \texttt{pair\_style uma \ldots devices 1} on every rank, so the
  graph-parallel world size equals the MPI world size.
  \item Combining multiple ranks with \texttt{devices${}>{}$1} (one rank
  forking several device workers) is a hard error: the two are alternative
  realizations of the same partition and mixing them double-shards the graph.
  \item Multi-rank operation additionally requires \texttt{precision double}.
  The multi-device runtime is FP64-only, so capacity cannot be traded for
  precision on this path, a limitation we return to in
  Sec.~\ref{sec:limitations}.
  \item Collectives use NCCL \cite{nccl} on NVIDIA GPUs (the engine is compiled for CUDA \cite{nickolls2008cuda}) and XCCL, native oneCCL \cite{oneccl}, on Intel GPUs (the engine is compiled with \texttt{icpx} for the SYCL C++ heterogeneous programming model \cite{sycl}), behind one internal peer interface offering \texttt{all\_gather\_nodes} and \texttt{all\_reduce\_sum}.
\end{itemize}

Throughout, $W$ counts \emph{devices as the runtime sees them}. An Intel node exposes 12 devices (GPU tiles) to the runtime, so $W{=}12$ per node and all Intel-node efficiencies are per tile. On NVIDIA nodes, $W$ equals the number of GPUs. Atom counts are rounded to the nearest thousand; ratios, per-atom costs, and the cube edge $N$ (in unit cells) use the exact cells.

\subsection{NPT virial path}
Isotropic \texttt{fix npt} at 300~K / 1~bar is the bulk default. The
pair style enables position+cell autograd when a barostat is present
and checkpointing is off. NVE/NVT do not need the strain gradient and
are used for vacuum nanoparticles, where padded-box NPT is
unphysical.

\subsection{Portability}
A thin device layer selects CUDA versus Intel XPU at build and run time, and the LAMMPS input, task head, and checkpoint-off NPT contract are identical on both GPU types. The claim we defend is stronger than input compatibility:
on one GPU type, results agree across device counts. Step-0 energies are identical to the last printed digit across every $W$ on NVIDIA GPUs (NCCL) and across every $W$ on Intel GPUs (XCCL). This does not claim that NVIDIA and Intel GPUs produce the same bits: the NVIDIA cell is a different random realization.
The ASE energy
and force deviations are in Sec.~\ref{sec:correctness}. For a deep equivariant
network evaluated in FP64 through a reordered all-reduce this is not automatic,
and it is what makes a result from either GPU type suitable for scientific use. Intel GPU numbers come from 12-tile Intel nodes, one to 32 nodes. NVIDIA GPU numbers come from 4$\times$A100-SXM4-40GB nodes and one 8$\times$H200 node.

\section{Implementation}

The engine is a C++ LAMMPS package plus a Python export of per-block
TorchScript modules (\texttt{uma-s-1p2}) \cite{paszke2019pytorch}. Export writes
\texttt{model\_traced.pt} and per-block/chunk artifacts; LAMMPS loads
them at \texttt{pair\_coeff}. Environment contract:

\begin{itemize}
  \item \texttt{UMA\_CHECKPOINT}: path to the FairChem \texttt{.pt}
  (export only).
  \item \texttt{UMA\_AC=off}, \texttt{UMA\_CKPT} unset for NPT.
  \item \texttt{UMA\_MP\_NATOMS}: atom count for the graph-parallel
  gather.
\end{itemize}

Activation checkpointing is \emph{off by default} and is enabled only by
opting in (\texttt{UMA\_AC=chunk|block|full} or \texttt{UMA\_CKPT=1}); when
either is set and a barostat is present, the pair style aborts with a diagnostic rather than integrating against an incorrect pressure. All NPT
results below therefore run the default. \texttt{UMA\_CHECKPOINT} names the
source checkpoint at export only, and \texttt{UMA\_MP\_NATOMS} selects the
artifact because partition offsets are baked into the TorchScript at trace
time, so an artifact is valid only for the $(N,W)$ it was traced at, which is a real constraint of tracing.

The graph is rebuilt inside every force evaluation. The LAMMPS neighbor list
is requested only to register the cutoff and is
never read; the engine constructs its own UMA graph from positions each step.
This is why the LAMMPS timing breakdown attributes 99.99\% of loop time to
\texttt{Pair} and $\approx$0.00\% to \texttt{Neigh}: graph construction,
network evaluation and the backward pass are all inside the pair call.
Double precision is required for the multi-device and NPT virial paths.

\section{Evaluation}\label{sec:eval}

We report the multi-node capacity and cost sweep (Sec.~\ref{sec:internode}), correctness, node-local capacity, and where the step goes. The demonstration that one checkpoint covers three chemistries, and the accuracy of those heating-ramp jumps, is Sec.~\ref{sec:melt}.

\subsection{Platforms}
All runs use \texttt{uma-s-1p2} in FP64. NaCl rocksalt at $8N^3$ atoms is the
capacity and parity workload; the heating ramps of Sec.~\ref{sec:melt} are the demonstration workload, not the capacity cells.
\begin{itemize}
  \item \textbf{NVIDIA, tight memory:} one 4$\times$A100-SXM4-40GB node.
  Checkpoint-off NPT ceiling and the node-boundary comparison.
  \item \textbf{NVIDIA, high memory:} one 8$\times$H200 node (141~GB/device).
  Largest checkpoint-off NPT cell, 47{,}000 atoms; the ten-step loop rate is 0.157~ns/day.
  \item \textbf{Intel:} 12-tile Intel nodes, one to 32 nodes.
  Checkpoint-off NaCl NPT ceiling 22{,}000 atoms on one node and 779{,}000 on 32; the ten-step loop rate is 0.10~ns/day on one node.
\end{itemize}

\subsection{Beyond one node: capacity and cost}\label{sec:internode}
Graph parallelism gathers the full node-feature tensor every block, so it
communicates $O(N)$ per layer and its per-device memory is $O(N)$ regardless
of world size.

We tested whether that is survivable across nodes on the
identical NaCl 4{,}000-atom cell, checkpoint-off NPT, over each machine's own network fabric.

Two comparisons isolate two effects. At equal world size (four A100s on one node versus two plus two, and twelve Intel tiles on one node versus six plus six), crossing the node boundary adds 2.5\% and 2.1\% per NPT step.
Energy, forces, and step-0 thermo agree across the split
($|\Delta E| = 2\times10^{-11}$~eV, $\max|\Delta\mathbf{F}| =
9\times10^{-16}$~eV/\AA; the A100 pressure is 10{,}978.496~bar on one
node and on two). Doubling the device count (four A100s versus eight, twelve Intel tiles versus twenty-four) does not shorten the step: 646~ms becomes 788~ms ($1.22\times$ longer) on A100, and 1{,}009~ms becomes 1{,}063~ms ($1.05\times$ longer) on the Intel node. On A100, about 80\% of that gap is per-rank GPU
compute that grows with world size; collectives are 14\%.
The step is compute on both machines: the collective is 2.0--4.5\% of the
A100 step and 0.5--2.1\% of the Intel step.
On Intel we separate communication from waiting by draining each rank's
queue and running a barrier before every timed collective. A host timer
around the all-reduce reads 665~ms on one node, but 646~ms of that is the
rank's own queued compute and 20~ms is rank skew; the collective itself is
5.5~ms. Crossing the node boundary at $W{=}12$ adds 16.5~ms of collective time
(1.6\% of the step). That time is host-observed: device-side collective
kernels take 3.5~ms at $W{=}12$ on one node and 3.8~ms at $W{=}24$ on two,
so most of the added time is not kernel time. TCP-socket two-node runs
inflate the Intel step $2.8\times$ and are not the numbers above.
The strain-gradient virial
adds 0.2~ms (A100) to 0.3~ms (Intel), so NPT and NVT differ by 1\% or
less at every layout. A second node is therefore a memory increment, and NPT is valid across it.

\begin{figure}[!htbp]
\centering
\includegraphics[width=\columnwidth]{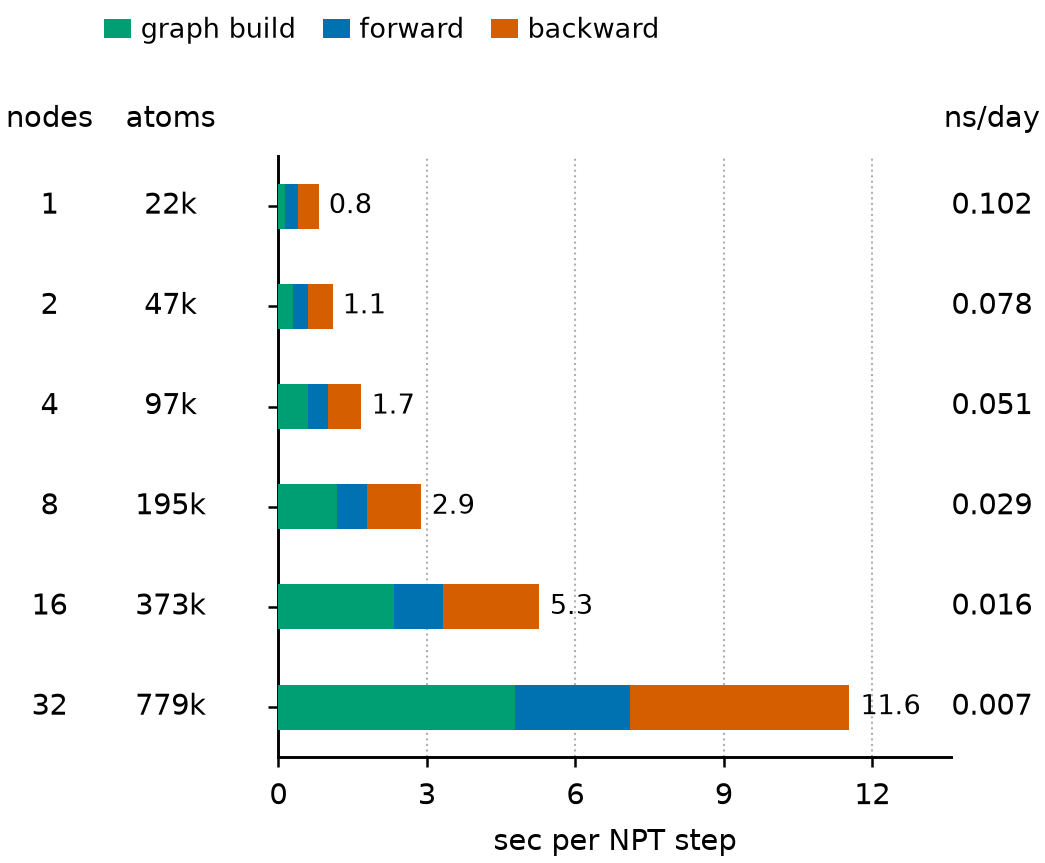}
\caption{Largest cubic NaCl cell that finished on 1--32 nodes of
Intel GPU nodes (12 tiles per node), ten NPT steps at 300~K and 1~bar, FP64,
checkpointing off. Bars are the per-step graph build, forward and backward
(rank 0, mean of the last nine force calls); the remaining $<$1.2\% is omitted.
Labels give the step in seconds, the atoms, and the ten-step loop rate.}
\label{fig:nodescale}
\end{figure}

\textbf{Capacity and speed across nodes.}
We ran cubic NaCl for ten NPT steps at 300~K and 1~bar on 1 to 32
12-tile Intel nodes ($W{=}12\times$nodes, FP64, checkpointing off, each machine's own network fabric), taking on each node count the largest cell that finished
(Fig.~\ref{fig:nodescale}). Capacity grows linearly with nodes, from 22{,}000
atoms to 779{,}000, 22--24 thousand per node throughout; the next size runs out
of device memory at every node count. The replicated $[N,C]$ tensor costs
9.2~kB per atom, 7.2~GB (11\% of one tile's memory) at 779{,}000 atoms, so
capacity must eventually lag the node count, but it has not at 32 nodes.
Speed does not follow. The step grows from 0.82~s (the single-node baseline of Sec.~\ref{sec:capacity}) to 11.6~s, so the ten-step
loop falls from 0.102 to 0.007~ns/day and weak-scaling efficiency from 74\% at two nodes to 7\% at 32. That is not the $\sim$2\% boundary cost: the 2\% is at equal world size on a fixed 4{,}000-atom cell, whereas weak scaling roughly doubles the cell with the devices. Forward and backward grow $8.6\times$ and
$10.5\times$. Graph build, a full-cell neighbor list that every rank repeats,
grows $38\times$, from 127~ms to 4.8~s: 15\% of the step on one node and 41\%
on 32. That is the lowest-effort remedy, since a shared or distributed
neighbor list needs no model surgery, though it leaves the forward and backward growth. Wall time per atom falls from 37~$\mu$s on one node to 14--15~$\mu$s from eight nodes on only because the cell grows $35\times$ while the step grows $14\times$; in device time, the cost per atom rises about $13\times$ from one node to 32 (37~$\mu$s on one node against 14.8~$\mu$s on 32). More devices therefore provide cell size, not a shorter step. Graph build alone is 5.8--6.2~$\mu$s per atom at every size, as expected of work every rank repeats for the whole cell. At 32 nodes the cell is 259~\AA\ across, so equal cubes for
384 ranks are 36~\AA\ wide against a 24~\AA\ receptive field: a 24~\AA\ halo would be $\sim$12 times the owned volume ($(84/36)^3\approx12.7$ for the padded region). One OpenMP thread per rank instead of the default thread count leaves capacity unchanged; the default is up to 12\% slower (16 nodes: ten-step loop 59.9 against 53.3~s).

For a foundation MLIP with a $\sim$24~\AA\ receptive field, one node is the unit of speed, and further nodes pool memory. The remaining multi-node axis is
\emph{data} parallelism over independent configurations, which needs no
cross-node collectives and scales trivially. Recovering multi-node
\emph{speed} for a single configuration would take a shared neighbor build
and, for the forward and backward growth, replacing the
per-block all-gather with a depth-$L$ halo exchange, which is model surgery on a released checkpoint that we do not attempt. That is a route to shorter
steps. It does not answer a measured communication bottleneck: the all-gather is $\sim$1\% of the 4{,}000-atom A100 step, the only size at which we split it, and we did not measure the share at multi-node cell sizes. A per-block exchange of one 6~\AA\ shell around 36~\AA\ cubes would carry $\sim$1.4 times the owned features per rank, against 384 times for the all-gather at 32 nodes (an estimate, not measured). Custom MLIPs of Sec.~\ref{sec:related} remain the appropriate comparison: their potentials are local, so their halo is thin and
their scaling is real.

\subsection{Correctness}\label{sec:correctness}
Against the ASE FairChem oracle on NaCl, Table~\ref{tab:ase-paths} lists the energy and per-atom force deviations of \texttt{pair\_style uma} (all atoms compared, not sampled).
An autograd (AG) versus finite-difference (FD) check on the energy at the
checkpoint-off $N{=}14$, $W{=}12$ cell gives
$\max|\mathrm{AG}-\mathrm{FD}| = 5.18\times10^{-7}$. The virial path is
gated behind the barostat, so NVT graph-parallel forces are bit-identical
to the non-virial build.

\textbf{Parity is against the reference implementation, not against density-functional theory (DFT).}
The numbers below are the deviation of this engine from the released UMA
model. They say nothing about how well UMA
reproduces first-principles forces, which is the model's question and is
addressed in \cite{wood2025uma}.

\textbf{Device-count consistency, checked on each GPU type.} The stronger correctness result
is that the answer does not depend on how many devices produced it.
On the checkpoint-off NPT ladders, step-0 and step-10 thermo agree across
every $W$ that fits: $W{=}1$--$12$ on one 12-tile Intel node,
$W{=}1,2,4$ on 4$\times$A100-40GB, and $W{=}1,2,4,8$ on H200. Table~\ref{tab:ase-paths} reports the ASE deviations
for \texttt{pair\_style uma} at $W{=}1$ and $W{=}4$ on A100 and on H200, and at $W{=}12$
on one 12-tile Intel node. The Intel ladder uses oneCCL; the CUDA ladders use NCCL.
Trajectories, not just single points, agree: multi-device step-10 energies
match the single-device trajectory. We obtain this by keeping the reduction order fixed with respect to $W$: the partition is a deterministic \texttt{tensor\_split} of a fixed atom ordering
and the per-shard reduction tree is built from that ordering, not from arrival order, so the all-reduce result does not depend on message timing.

\textbf{Two LAMMPS paths versus ASE.}
The numbers above compare \texttt{pair\_style uma} to the FairChem
graph-parallel Python driver. FairChem also ships a LAMMPS path
(\texttt{fairchem-lammps}) that injects forces through \texttt{fix external} rather than a pair style \cite{fairchemlammps,thompson2022lammps}.
Table~\ref{tab:ase-paths} holds both paths to the same ASE
\texttt{FAIRChemCalculator} baseline on rattled NaCl rocksalt (FP64, task
\texttt{omat}; energy and every per-atom force, not a subsample) across
$N{=}6$ ($W{=}1$), $N{=}10$ ($W{=}4$), and $N{=}14$ ($W{=}12$) on NVIDIA and Intel GPUs.
The result is a clear separation that holds at every cell and on
every device: \texttt{pair\_style uma} tracks the reference at the FP64 floor,
$|\Delta E|\sim10^{-10}$--$10^{-8}$~eV and
$\max|\Delta\mathbf{F}|\sim10^{-14}$~eV/\AA, while the wrap's deviations lie four orders of magnitude higher in energy and eight in force,
$10^{-6}$--$10^{-5}$~eV and $10^{-6}$--$10^{-5}$~eV/\AA\
(Table~\ref{tab:ase-paths}). The gap is the input precision, not the model:
the wrap's \texttt{fix external} callback builds positions and the cell as FP32 tensors and upcasts them for the FP64 model, so the coordinates, not the UMA weights, set that residual.
FP32 unit roundoff is $\varepsilon_{\mathrm{FP32}}\approx1.2\times10^{-7}$;
the measured wrap forces lie $\sim$50--100$\times$ above that unit roundoff.

This is a numerical-precision distinction, and it is the one GPU MD engines are built around. GPU codes expose precision modes:
all-double (DPDP), single-precision math with double-precision or 64-bit fixed-point accumulation
(mixed, SPDP/SPFP), and all-single (SPSP)
\cite{gotz2012amber,legrand2013spfp}. The field's consensus is that all-single is unsafe for production, because its rounding error accumulates over a long trajectory. AMBER reports SPSP producing unphysical trajectories in generalized-Born simulations and recommends mixed precision as the floor \cite{gotz2012amber}; GROMACS runs its GPU path in mixed precision \cite{kutzner2015gromacs}; and the HALMD single-precision GPU integrator drifts about $0.1\%$ in total energy per $10^{7}$ steps \cite{colberg2011glassy}.
The mixed-precision recipe is not available for UMA. Its forces are the
autograd gradient of the energy and inherit the precision of the network, so
the released model runs in single or double precision, not in between.
\texttt{pair\_style uma} takes the double-precision side of that choice, the
whole force and energy path in FP64 (DPDP), a mode usually reserved
for CPUs because GPU FP64 throughput is lower; graph parallelism is what makes
it affordable here. Against the FP64 ASE baseline the \texttt{pair\_style uma} force residual is
$10^{-14}$~eV/\AA\ and the wrap's is $10^{-6}$--$10^{-5}$~eV/\AA, with the
energy gap four orders wide. The wrap's model also runs in FP64, so that
residual comes from its single-precision coordinate input; it is identical on
A100 and H200. We did not run the wrap's default single-precision model, whose
residual would be larger. The comparison is fidelity to the released model,
not accuracy against first-principles forces, and we do not compare long-run
trajectories between the two.
\begin{table}[!htbp]
\caption{ASE FairChem as baseline. NaCl rocksalt, FP64, all atoms.
$N{=}6$ is 2{,}000 atoms ($W{=}1$); $N{=}10$ is 8{,}000 atoms ($W{=}4$);
$N{=}14$ is 22{,}000 atoms ($W{=}12$).
A100 is the 40~GB model.
$\Delta$ is LAMMPS minus ASE.
Path \texttt{uma} is \texttt{pair\_style uma}; wrap is FairChem's \texttt{fix external}.
$|\Delta E|$ in eV; $\max|\Delta\mathbf{F}|$ in eV/\AA.}
\begin{center}
\footnotesize
\setlength{\tabcolsep}{3pt}
\begin{tabular}{@{}llrrr@{}}
\toprule
\textbf{Node} & \textbf{Path} & \textbf{$N$} &
\textbf{$|\Delta E|$} & \textbf{$\max|\Delta F|$} \\
\midrule
1$\times$A100 & \texttt{uma} & 6 & $1.3{\times}10^{-10}$ & $1.9{\times}10^{-14}$ \\
1$\times$A100 & wrap & 6 & $3.0{\times}10^{-6}$ & $6.4{\times}10^{-6}$ \\
1$\times$H200 & \texttt{uma} & 6 & $1.3{\times}10^{-10}$ & $2.0{\times}10^{-14}$ \\
1$\times$H200 & wrap & 6 & $3.0{\times}10^{-6}$ & $6.4{\times}10^{-6}$ \\
4$\times$A100 & \texttt{uma} & 10 & $4.0{\times}10^{-9}$ & $2.9{\times}10^{-14}$ \\
4$\times$A100 & wrap & 10 & $2.7{\times}10^{-5}$ & $1.1{\times}10^{-5}$ \\
4$\times$H200 & \texttt{uma} & 10 & $4.0{\times}10^{-9}$ & $2.9{\times}10^{-14}$ \\
4$\times$H200 & wrap & 10 & $2.7{\times}10^{-5}$ & $1.1{\times}10^{-5}$ \\
12-tile Intel & \texttt{uma} & 14 & $3.3{\times}10^{-8}$ & $8.8{\times}10^{-14}$ \\
12-tile Intel & wrap & 14 & $1.5{\times}10^{-5}$ & $1.9{\times}10^{-5}$ \\
\bottomrule
\end{tabular}
\label{tab:ase-paths}
\end{center}
\end{table}

\subsection{NPT step cost and capacity}\label{sec:capacity}
\begin{figure}[!htbp]
\centering
\includegraphics[width=\columnwidth]{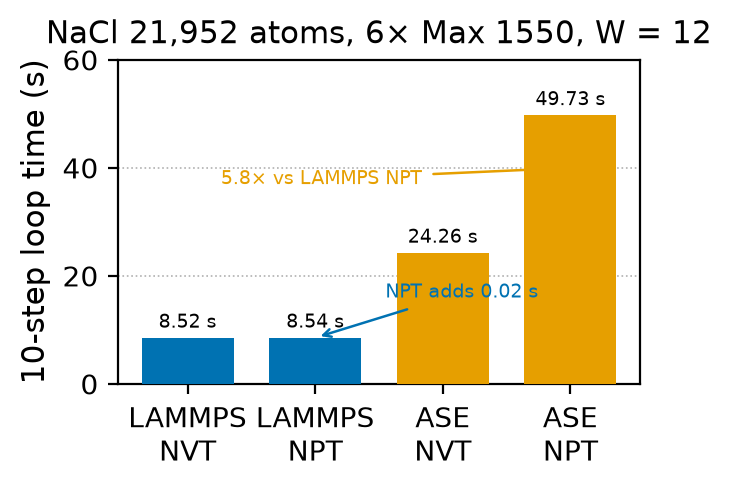}
\caption{Ten-step loop time at 22{,}000 NaCl atoms on one 12-tile Intel node, $W{=}12$, checkpointing off.
LAMMPS NPT is 8.54~s and NVT is 8.52~s.
The Python driver incurs 24.26~s for NVT and 49.73~s for NPT because it
recomputes the stress. Integrators differ (Nos\'{e}--Hoover versus
Berendsen-style cell coupling), so the driver bars are cost per step.}
\label{fig:barostat}
\end{figure}

\textbf{Step cost.} At the largest node-local NPT size on one
12-tile Intel node (22{,}000 atoms), ten MD steps cost 8.52~s under NVT
and 8.54~s under NPT: the strain gradient and its all-reduce add 0.2\%. A
device-synchronized phase profile of these runs puts the
virial at 0.2--1.2~ms and the NPT$-$NVT step difference at 0.6--1.9~ms.
Backward does not grow. The virial is already available from the same
backward pass that produces forces, and reducing six additional scalars
across the world is not measurable against an $O(N)$ node-feature gather.
The Python driver on the identical
geometry incurs 24.26~s for NVT and 49.73~s for NPT, because its barostat
recomputes the stress (Fig.~\ref{fig:barostat}). Those times are $2.9\times$
and $5.8\times$ the engine, and the integrators are not identical
(Nos\'{e}--Hoover here, Berendsen-style cell coupling in the driver), so
this is a cost-per-step comparison, not a trajectory one.
We also timed FairChem's Python-wrapped LAMMPS path (\texttt{fairchem-lammps}, \texttt{fix external}) \cite{fairchemlammps} on a smaller cell: NaCl $N{=}6$ (2{,}000 atoms) on one A100, ten NPT steps. There, \texttt{pair\_style uma} loops in 4.63~s (pair 99.98\%), while the wrap needs 2.96~s to load, 7.36~s for a first ten steps that include LAMMPS construction, and 3.90~s for a second ten steps. On one H200, the same cell loops in 2.19~s with \texttt{pair\_style uma} (pair 99.96\%), while the wrap needs 2.60~s to load, 5.90~s for a first ten steps, and 1.60~s for a second ten steps. That wrap puts the force in LAMMPS Modify (Pair 0\%),
so it is not the same timer split. On this $W{=}1$ cell the wrap is not
slower. On the 22{,}000-atom $W{=}12$ cell the wrap itself loops ten NPT
steps in 8.07~s (0.107~ns/day) against 8.54~s for \texttt{pair\_style uma}. \texttt{pair\_style uma} does not reduce wall time relative to FairChem's LAMMPS wrap. It instead keeps the force and energy path in double precision and forms the strain-gradient virial inside the pair call. \texttt{pair\_style uma} fits this cell only at $W{=}12$. The wrap also produces forces at $W{=}8$ and $W{=}10$, where \texttt{pair\_style uma} runs out of memory, so on the Intel node the wrap fits this cell at more layouts; we claim no capacity advantage over it. Table~\ref{tab:ase-paths} lists the energy and per-atom force deviations of both LAMMPS paths against ASE, including the 8{,}000-atom $W{=}4$ cells on 4$\times$A100-40GB and 4$\times$H200 and the 22{,}000-atom $W{=}12$ cell on 12-tile Intel.

\textbf{Capacity.} With checkpointing off, retained activations set the cell
size. On one
8$\times$H200 node (141~GB/device) the ceiling is
\textbf{47{,}000 NaCl atoms} ($N{=}18$), with $N{=}19$ (55{,}000 atoms)
exhausting device memory at artifact export (137.7~GiB committed when a
512~MiB allocation failed); the cell that fits finishes at 105.75~GiB
allocated. That is the largest strain-gradient NPT cell we report,
$5.8\times$ the tight-memory A100 ceiling and $2.1\times$ the
12-tile Intel ceiling. On one 4$\times$A100-40GB node, 8{,}000 NaCl
atoms finish a ten-step 300~K loop and run out of memory at 11{,}000. Export traces a
fixed graph, so a longer or hotter NPT can exceed that cap at the same atom
count; a production cell is therefore smaller: 8{,}000 NaCl atoms exceed
the cap on the first step of an 800~K ramp, while 6{,}000 complete the
100~ps ramp. Neighbor density depends on chemistry, so atom count is not the
right unit. CUDA managed memory thrashed on this path, and edge chunking
does not raise NPT capacity, because retained activations bind it.

\begin{figure}[!htbp]
\centering
\includegraphics[width=\columnwidth]{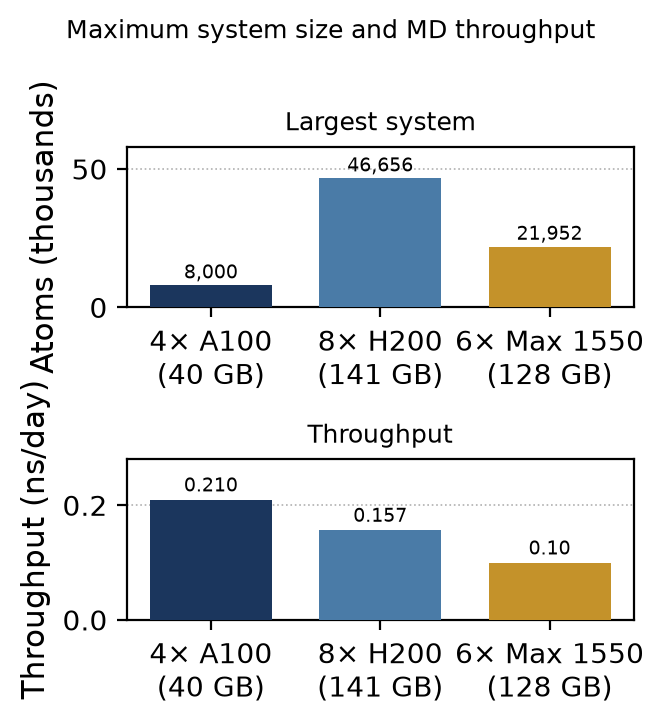}
\caption{Checkpoint-off NaCl NPT ceilings, one node each.
Top: largest cell that fits, 8{,}000 atoms on 4$\times$A100-40GB,
47{,}000 on 8$\times$H200, and 22{,}000 on one 12-tile Intel node.
Bottom: ten-step loop rate at that same cell. The A100 bar is
the highest rate because the cell is the smallest; the three bars are not a
same-size comparison, and each rate is a setup-influenced 10-step loop.
A longer NPT can require a smaller cell than the ten-step ceiling.}
\label{fig:machine-scale}
\end{figure}

\textbf{Throughput.} At full-node ceilings the ten-step loop rates, which
are not production throughputs, are
0.210~ns/day on one 4$\times$A100-40GB node (8{,}000 atoms),
0.10~ns/day on one 12-tile Intel node (22{,}000 atoms), and
0.157~ns/day on 8$\times$H200 (47{,}000 atoms)
(Fig.~\ref{fig:machine-scale}), with $>$99.9\% of loop time in the pair call.
The same H200 cell over 10{,}000 steps at 300~K is 0.149~ns/day.
The production runs are the nine finished 100~ps ramps of Sec.~\ref{sec:melt},
on cells smaller than these ten-step ceilings. A wall-day at the ten-step rates is tens of
picoseconds, which is sufficient for lattice parameters, radial distribution
functions (RDFs) and coordination, and not for diffusion or long-time
kinetics.

\subsection{Device count governs capacity, not throughput}
Within a node, device count $W$ sets the cell that fits, not the loop time.
The $[N,C]$ node tensor is replicated and the backward pass carries an
in-graph all-reduce (Sec.~\ref{sec:roofline}), so adding tiles raises the
checkpoint-off NPT ceiling rather than shortening the step; retained
activations set the per-device ceiling, and we choose $W$ to fit the cell.

The resulting node-local NPT ceilings are these. On one 4$\times$A100-40GB
node the one-GPU cell is NaCl $N{=}6$ (2{,}000 atoms). On one
12-tile Intel node the node ceiling is $N{=}14$ (22{,}000 atoms), which
fits only at $W{=}12$ (step-0 energy $-74141.205$~eV, pressure
11{,}032.4~bar); the same atoms at $W{=}8$ run out of memory. On H200 a single device holds $N{=}9$ (6{,}000 atoms) and the full 8$\times$H200 node reaches 47{,}000 atoms. Relative to one device, the regime of the ASE calculator, graph parallelism therefore raises the ceiling $4.6\times$ on A100 (2{,}000 to 8{,}000 atoms) and $8.0\times$ on H200. Step-0 and step-10 thermo agree across every $W$ that fits, with both NCCL and oneCCL collectives, so device count changes capacity
without changing the trajectory.

Because the step is replicated, adding tiles does not shorten the loop on small cells (one tile is fastest at $N{=}6$). It begins to help once the cell is large enough to amortize the collective: on the 6{,}000-atom H200 cell, four devices cut the loop to 2.76~s from 5.27~s at $W{=}1$,
which is $1.91\times$, 48\% of ideal. Eight devices on that cell take
4.12~s, slower than four. 
The cost of each extra tile is per-rank work that grows with the cell, not the
barostat; Sec.~\ref{sec:roofline} gives the phase split.

\subsection{Where the step goes}\label{sec:roofline}
On the 12-tile Intel node the profiled checkpoint-off cell is
$N{=}14$, $W{=}12$. Its ten-step loop is 8.47~s and the steady step is 820~ms
(mean of the last nine force calls): graph build 127~ms, of which 118~ms is
the vesin neighbor list; forward 270~ms, which holds the node all-gathers; and
backward 422~ms, which holds the in-graph all-reduces. The drain-and-barrier
split of Sec.~\ref{sec:internode} measures the collective share on the
4{,}000-atom cell: 0.5--2.1\% of the step on the Intel node and 2.0--4.5\% on A100.
On that cell the step is compute, not communication. We did not repeat the split on the 22{,}000-atom cell or on the multi-node cells, so we quote no collective share for them. The post-graph force and virial sums are one $[N,3]$ FP64 tensor and
one $[3,3]$ cell gradient, $24N$ and 72~bytes (about 0.53~MB at 22{,}000
atoms). These sums take about 1~ms on the Intel node and under 1~ms on A100 at every layout; the force all-reduce takes 0.12~ms on one 4$\times$A100-40GB cell (8{,}000 atoms) and 0.21~ms on one 8$\times$H200 cell (47{,}000 atoms). They do not account for the two-node cost (Sec.~\ref{sec:internode}).

A FairChem graph-parallel flop counter (PyTorch \texttt{FlopCounterMode}) on
the 8{,}000-atom A100 cell ($N{=}10$, $W{=}4$)
counts $2.006\times10^{12}$ FLOPs per rank for one forward-plus-backward
predict. Dividing by the \texttt{pair\_style uma} last step on that cell, 408~ms, gives $4.92\times10^{12}$~FLOP/s per GPU, 51\% of the documented
9.7~TFLOP/s FP64 peak of an A100-40GB SXM \cite{a100datasheet}. That quotient is a labeled lower
bound on the Python graph-parallel predictor, not a hardware counter in
\texttt{pair\_style uma}: the 1-GPU profiler returns 0 counted FLOPs on
$N{=}6$ because UMA's custom kernels are missing from the flop table.
The same counter on the 47{,}000-atom H200 cell ($N{=}18$, $W{=}8$)
counts $5.847\times10^{12}$ FLOPs per rank (node-sum
$4.678\times10^{13}$). The \texttt{pair\_style uma} last step on that cell is 548~ms, so the
quotient is $1.07\times10^{13}$~FLOP/s per GPU, 31\% of the documented
34~TFLOP/s FP64 peak of an H200 SXM \cite{h200specs}, labeled the same way.
The two cells are not the same size, so the count is not transferred to the
12-tile Intel ceiling.

The demonstration workload (100~ps NPT heating ramps of NaCl, equiatomic Pt--Ru, and equiatomic (HfNbTaTiZr)N from one \texttt{omat} checkpoint) is the subject of Sec.~\ref{sec:melt}. Those ramps use the checkpoint-off NPT path whose
capacity and cost are measured above; they are not melting-point
determinations. Section~\ref{sec:melt-accuracy} states what the jumps are
and what the experimental marks are.

\section{Heating ramps of three chemistries}\label{sec:melt}
One \texttt{omat} checkpoint drives NPT MD for three chemistries that have no
shared classical potential (a salt, a binary alloy, and a five-cation nitride) with no refit. That
single-checkpoint coverage is the systems result; the heating ramps are how we
exercise the NPT path of Sec.~\ref{sec:eval}. This section reports the protocol
and nine jumps. Section~\ref{sec:melt-accuracy} is the accuracy
discussion: the jumps are not melting points.

\subsection{Protocol and observed jumps}
\begin{figure}[!htbp]
\centering
\includegraphics[width=\columnwidth]{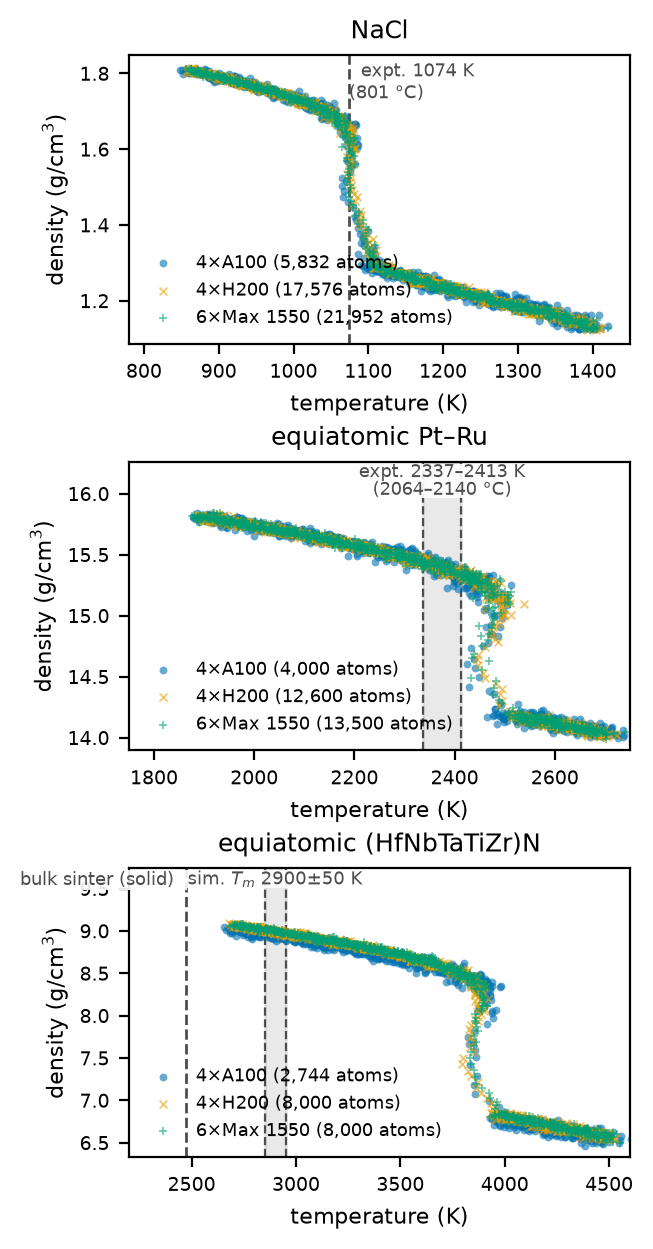}
\caption{Density versus temperature after the first 10~ps of each production
NPT ramp, one \texttt{omat} checkpoint. Dashed marks are experimental or
literature context: NaCl melting point \cite{crc2022}; Pt--Ru
solidus--liquidus at 50~at.\% Ru \cite{cornish2007ptru}. On the (HfNbTaTiZr)N
panel the shaded band is the simulated melting temperature $2900\pm50$~K
computed for this same composition by a heating/cooling hysteresis method
\cite{baidyshev2024hecn},
and the dashed line at 2473~K is the bulk spark-plasma-sintering densification
temperature, a solid-state lower bound \cite{dippo2020bulkhen}; the nitride has no directly measured melting point, and even UN needs N$_2$ overpressure to melt \cite{carvajal2014un}. Jumps are where
a defect-free ramp fails, not melting-point estimates. See
Sec.~\ref{sec:melt-accuracy}.}
\label{fig:melt-tdens}
\end{figure}

The same \texttt{omat} checkpoint heats three crystals on one isotropic NPT
path: NaCl, equiatomic Pt--Ru, and equiatomic rocksalt
(Hf$_{0.2}$Nb$_{0.2}$Ta$_{0.2}$Ti$_{0.2}$Zr$_{0.2}$)N. The Zhou--Johnson--Wadley embedded-atom table covers 16 metals and no salt \cite{zhou2004eam}. The nitride cations are taken from special-quasirandom
occupations of that composition \cite{zunger1990sqs}. Each production run is
100~ps (100{,}000 steps of 1~fs) at 1~bar, with thermostat and barostat damping
of 0.1~ps and 1.0~ps. Checkpointing stays off, because the barostat uses the
strain-gradient virial. The ramps are 6~K/ps for NaCl, 9~K/ps for Pt--Ru, and
20~K/ps for the nitride. The signature is the jump in potential energy per
atom and in density. Figure~\ref{fig:melt-tdens} is density versus temperature
for the nine ramps.

Table~\ref{tab:jumps} lists the nine jumps: the density drop and the temperature at which it occurs, which stays within a window of at most 120~K over the drop.

\begin{table}[!htbp]
\caption{Density drop and jump temperature of the nine 100~ps NPT ramps.}
\label{tab:jumps}
\begin{center}
\footnotesize
\setlength{\tabcolsep}{3pt}
\begin{tabular}{@{}llrrr@{}}
\toprule
\textbf{System} & \textbf{Node} & \textbf{Atoms} & \textbf{$\rho$ (g/cm$^3$)} & \textbf{Jump (K)} \\
\midrule
NaCl & 4$\times$A100 & 6{,}000 & 1.65$\to$1.28 & 1070 \\
NaCl & 4$\times$H200 & 18{,}000 & 1.64$\to$1.31 & 1090 \\
NaCl & 12-tile Intel & 22{,}000 & 1.63$\to$1.31 & 1090 \\
Pt--Ru & 4$\times$A100 & 4{,}000 & 15.2$\to$14.2 & 2470 \\
Pt--Ru & 4$\times$H200 & 13{,}000 & 15.0$\to$14.2 & 2500 \\
Pt--Ru & 12-tile Intel & 14{,}000 & 15.2$\to$14.2 & 2480 \\
(HfNbTaTiZr)N & 4$\times$A100 & 3{,}000 & 7.82$\to$6.99 & 3870 \\
(HfNbTaTiZr)N & 4$\times$H200 & 8{,}000 & 7.69$\to$7.03 & 3870 \\
(HfNbTaTiZr)N & 12-tile Intel & 8{,}000 & 7.81$\to$7.03 & 3870 \\
\bottomrule
\end{tabular}
\end{center}
\end{table}

The cells in Fig.~\ref{fig:melt-tdens} are the production ramps. Neighbor density depends on chemistry, so each ramp uses a cube that stays under the export cap as the neighbor list grows: Pt--Ru $15^3$ (14{,}000 atoms) aborted at step 0 on 4$\times$H200 because two ranks held about 197{,}000 edges each, so that node ran a $15{\times}15{\times}14$ cell (13{,}000 atoms). Three ramps stopped once on a filesystem write and were rerun to completion. The A100 cubes are NaCl $9^3$, Pt--Ru $10^3$, and nitride $7^3$; the Intel cubes are NaCl $14^3$, nitride $10^3$, and Pt--Ru $15^3$. All nine 100~ps ramps finished. The nitride cells use special-quasirandom equiatomic
occupations of Hf, Nb, Ta, Ti, and Zr on the cation sublattice, built from
twelve tiles at the rocksalt (B1) lattice constant $a=4.4845$~\AA, with no new
fit and no new tiling.

\subsection{Accuracy of the jumps}\label{sec:melt-accuracy}
The nine jumps are where a defect-free NPT ramp fails. They are not melting
points of the crystals, of the potential, or of experiment. Four facts about
the protocol set that number, and the experimental marks are context, not a
score.

A periodic, defect-free crystal with no solid--liquid interface superheats
\cite{luo2003superheat}. Homogeneous nucleation has to create the liquid
inside the bulk; the barrier is large on molecular-dynamics timescales, so
the solid survives above the equilibrium melting temperature of the
potential. The accepted way to measure that equilibrium temperature is a
two-phase coexistence simulation \cite{morris2002coexistence}. We run
neither coexistence nor a cooling branch. The ramps are fast (6, 9, and 20~K/ps), and the
heating \emph{rate} extra-superheats the solid. The cells have no free
surface, grain boundary, dislocation, or vacancy to nucleate
against, and they are small (3{,}000--22{,}000 atoms); homogeneous
nucleation scales with volume, so a smaller cube typically fails
later. Each reported jump is the density drop over about 1~ps
while temperature is still moving, not a free-energy crossing.

We have no coexistence $T_m$ for \texttt{uma-s-1p2}/\texttt{omat} \cite{wood2025uma}, so
we cannot split ramp superheat from model-versus-experiment error.

The experimental marks in Fig.~\ref{fig:melt-tdens} are therefore not an
accuracy score for the engine. NaCl melts at 1074~K (801~$^\circ$C) \cite{crc2022}. The three NaCl jumps
(1070--1090~K) lie near that number by proximity only. Platinum melts at 2041~K (1768~$^\circ$C) and
ruthenium at 2607~K (2334~$^\circ$C) \cite{crc2022}. At 50~at.\% Ru the Massalski
experimental diagram, reproduced as Fig.~6(b) of Cornish
\textit{et al.}~\cite{cornish2007ptru}, is a solidus--liquidus lens: about
2337~K (2064~$^\circ$C) solidus and 2413~K (2140~$^\circ$C) liquidus, read from the published axes. The
1800 to 2700~K ramp brackets the interval; all three Pt--Ru jumps
(2470--2500~K) lie above the liquidus, which is the expected sign for a
superheated homogeneous failure.

Equiatomic rocksalt (HfNbTaTiZr)N has no directly measured melting point.
Transition-metal nitrides lose N$_2$ and decompose before they melt
congruently at ambient pressure; even uranium nitride needs 0.25~MPa of N$_2$ to melt at
$3120\pm30$~K \cite{carvajal2014un}. The MD cell is closed, stoichiometric, and held at 1~bar,
with no N-vacancy population, so a defect-free ramp cannot reproduce a nitride melting experiment even in principle.

The literature that does bound this composition is consistent with the jump
being a high superheated failure, not a $T_m$. Bulk (HfNbTaTiZr)N survives
spark-plasma-sintering densification at 2473~K (2200~$^\circ$C), a solid-state
lower bound \cite{dippo2020bulkhen}. A heating/cooling hysteresis simulation of
this same five-cation nitride places its melting temperature at $2900\pm50$~K
\cite{baidyshev2024hecn}, and the equiatomic rule-of-mixtures of the five
binary end-members (HfN, ZrN, TiN, TaN, NbN) gives $\approx$3190~K
\cite{crc2022}. All three MD jumps ($\approx$3870~K) lie above every one of
these anchors: above the hysteresis-derived $T_m$ by about 1000~K and above
the binary average by about 700~K. That ordering is the expected sign of homogeneous failure in a small, defect-free, fast ramp; it is not a melting-point determination.

\section{Limitations}\label{sec:limitations}

Multi-node speed-up of a single configuration is not delivered, and
Sec.~\ref{sec:internode} is why: per-rank work is $O(N)$ in the global atom
count, so a second node adds memory but not speed. Multi-node \emph{data}
parallelism over independent configurations is available and untested at
scale. Checkpointing stays off on the NPT path because we have not validated the cell gradient through our recompute functions, so the cell size is set by retained activation memory. Because tracing bakes
partition offsets, $N$ and $W$ are fixed at export: no dynamic load
balancing, no resizing mid-run. The graph is rebuilt every
step with no reuse, correct but wasteful for small displacements.
Across nodes the step grows with the cell: every rank rebuilds the full-cell
neighbor graph, which is 15\% of the step on one node and 41\% on 32
(Sec.~\ref{sec:internode}), and forward and backward grow with $N$ as well.
Each row of that sweep is one run on rank 0, and the 2\% boundary cost and the 1\% NPT-versus-NVT difference are single measurements per layout, with no run-to-run variance. The collective split exists only at 4{,}000 atoms, so the compute-bound reading does not extend to the multi-node cells.
The multi-device path requires \texttt{precision double}, so precision cannot currently be traded for capacity. The released model offers only single or double precision, so raising the NPT ceiling that way would mean validating an FP32 path against the FP64 one, which we have not done; it is future work. We also make no per-atom speed comparison with the spatially decomposed MLIPs of Sec.~\ref{sec:related}. Code and scripts will be released at a later date. Wall-day trajectories are tens of picoseconds: enough
for lattice constants, RDFs and coordination, not for diffusive annealing or
catalytic turnover. The heating-ramp jumps are not melting points; the
protocol and experimental marks are in Sec.~\ref{sec:melt-accuracy}.

\section{Conclusion}
Foundation MLIPs occupy a different point in the parallel design space from
the custom-made potentials that have reached $10^{8}$--$10^{10}$ atoms. Their
generality comes from depth and global couplings, which force graph
parallelism. That makes one node the unit of speed, which we support by
measurement: crossing a node boundary at equal world size adds about 2\% to the
NPT step on a 4{,}000-atom cell, and doubling devices does not shorten it. Three
results hold within a node. The strain-gradient virial adds 0.2\% per step
against NVT. The results do not depend on the device count or on whether NCCL or oneCCL provides the collectives, and the FP64 path agrees with ASE to $\sim10^{-14}$~eV/\AA, eight orders of magnitude below the residual caused by the wrap's FP32 coordinates. In our implementation
NPT costs capacity rather than time: ten-step NPT loops hold 8{,}000, 22{,}000, and 47{,}000 NaCl atoms on the three node types (the Intel-node figure is 20 times below the checkpoint-on NVT ceiling), and capacity reaches 779{,}000
atoms on 32 Intel nodes at 0.007~ns/day. The pair style runs on NVIDIA and Intel GPUs from one source tree. It trades speed per atom for one checkpoint; with it, NPT ramps of a salt, an alloy, and a nitride run with no refit. Two directions follow: carrying the cell
gradient through checkpoint recompute to close the 20-fold gap, and recovering
multi-node speed through a shared neighbor build and, further out, a depth-$L$
halo exchange.

\section*{Acknowledgments}
This work was supported by Laboratory Directed Research and Development (LDRD) funding from Argonne National Laboratory, provided by the Director, Office of Science, of the U.S.\ Department of Energy under Contract No.\ DE-AC02-06CH11357. This research used resources of the Argonne Leadership Computing Facility, which is a U.S.\ Department of Energy Office of Science User Facility operated under contract DE-AC02-06CH11357. This research used the Delta advanced computing and data resource which is supported by the National Science Foundation (award OAC 2005572) and the State of Illinois. Delta is a joint effort of the University of Illinois Urbana-Champaign and its National Center for Supercomputing Applications.

We used Claude Sonnet 5.5 \cite{claudesonnet} to support article writing and editing. The authors reviewed the text and are responsible for its content.

\FloatBarrier

\end{document}